\documentclass{iopart}
\usepackage{iopams}

\begin{document}

\title[Quantum field algebra]{A quantum field algebra}

\author{Christian Brouder\dag\footnote[3]{(christian.brouder@lmcp.jussieu.fr)}}

\address{\dag\ Laboratoire de Min\'eralogie-Cristallographie, CNRS UMR7590,
 Universit\'es Paris 6 et 7, IPGP, 4 place Jussieu,
  75252 Paris Cedex 05, France}

\begin{abstract}
The Laplace Hopf algebra created by Rota and coll. is 
generalized to provide an algebraic tool for
combinatorial problems of quantum field theory. 
This framework encompasses commutation relations,
normal products,
time-ordered products and renormalisation.
It considers the operator product and the time-ordered
product as deformations of the normal product.
In particular, it gives an algebraic meaning to 
Wick's theorem and it extends the concept of Laplace
pairing to prove that the renormalised
time-ordered product is an associative deformation
of the normal product
involving an infinite number of parameters.
The parameters themselves form a group: the renormalisation group,
which acts on the product instead of on the algebra.
\end{abstract}

\today


\submitto{\JPA}

\maketitle
\newcommand{\Id}{{\mathrm{Id}}}
\newcommand{\sign}{{\mathrm{sign}}}
\newcommand{\perm}{{\mathrm{perm}}}
\newcommand{\ee}{{\mathrm{e}}}
\newcommand{\barT}{\bar{T}}
\newcommand{\bart}{\bar{t}}
\newcommand{\barG}{\bar{G}}
\newcommand{\barc}{\bar c}
\newcommand{\barV}{\bar V}
\newcommand{\Lcal}{{\cal L}}
\newcommand{\Scal}{{\cal S}}
\newcommand{\Acal}{{\cal A}}
\newcommand{\Xcal}{{\cal X}}
\newcommand{\Z}{{\zeta}}
\newcommand{\ZZ}{{Z}}
\newcommand{\Jcal}{{\cal J}}
\newcommand{\antip}{{s}}
\newcommand{\barcirc}{{\bar \circ}}
\newcommand{\rlaplace}[2]{({\overline{#1|#2}})}
\newcommand{\binom}[2]{\left(\begin{array}{c}#1\\#2\end{array}\right)}

\section{Introduction}
Quantum field theory is one of the most powerful methods of
contemporary physics. It exhibits exciting analytical
and combinatorial problems that are tightly intermingled.
The purpose of this paper is to develop an algebraic
framework which simplifies the solution of its
combinatorial problems. This framework encompasses the
main concepts used in the practical calculation
of observable quantities: normal products,
time-ordered products and renormalisation. 
With these three tools, one can calculate the S-matrix 
and the Green functions, i.e. about everything that
can be compared with experiments.
The analytical difficulties will be avoided by working 
in a finite-dimensional vector space instead of with 
field operators (i.e.  operator-valued distributions \cite{Wightman}).
The algebraic framework is an extension of the ``Laplace Hopf algebras'' 
created and elaborated by Rota and coll. 
\cite{Doubilet,Grosshans}. A Laplace Hopf algebra is a
Hopf algebra equipped with a bilinear form
called the ``Laplace pairing''. 
With this algebra, Rota and coll. were able to
transform intricate combinatorial problems into 
clear algebraic manipulations
\cite{RotaStein86,RotaStein89,RotaStein90_1,RotaStein90_2,RotaStein94}.

In quantum field theory, the commutator of two field operators
is not an operator but a scalar function. Therefore,
all products of field operators can be written as a linear 
combination of symmetric (i.e. normal) products of operators. 
We denote by $S(V)$ the vector space spanned by
all normal products of operators.
Similarly, the difference between the time-ordered product
of two operators and the normal product of these
operators is a scalar function. Thus, all time-ordered
products of operators belong also to $S(V)$.
In other words, quantum field theory can be entirely
discussed in the space $S(V)$ of normal products.
Working in $S(V)$ has several advantages, already
noted by Houriet and Kind \cite{Houriet}, and later
stressed by Wick \cite{Wick} and by Gupta
\cite{Gupta}: $S(V)$ is equipped with a commutative
product (the symmetric product) and the
expectation value of an element $u$ of $S(V)$ over the vacuum
is zero if $u$ has no scalar component.

Fauser discovered recently \cite{Fauser} that the time-ordered product
is an instance of the  ``circle product'', which is 
a deformation of the symmetric product of $S(V)$ constructed from
the Laplace pairing. In particular, Wick's
theorem becomes a very simple algebraic expression.
Moreover, we prove that the operator product is also
an instance of the circle product derived from
a different Laplace pairing. Thus, the bare quantities
of quantum field theory can be obtained from circle products.
However, renormalisation does not fit into 
Rota and Stein's original framework.
Renormalisation was created in 1949
by Dyson \cite{Dyson}, and its algebraic meaning
remained mysterious until Kreimer discovered
that it is ruled by a Hopf algebra \cite{Kreimer98}.
Renormalisation was then refined by Connes and
Kreimer \cite{CKI,CKII}. There are several ways
to see renormalisation. In the standard approach,
an analytic expression is renormalised by 
adding counterterms to the Lagrangian. 
Epstein and Glaser proposed an alternative point 
of view, where renormalisation is seen as 
a modification of the time-ordered product
\cite{Epstein}. This approach was elaborated
into its most powerful form by Pinter 
\cite{Pinter,PinterHopf}. 
It turns out that her work can be used to
define a deformation of the circle product
through a modified Laplace pairing, which satisfies 
axioms weaker than those of  Rota's Laplace pairing,
but enables us to define an associative
product, the renormalised circle product. This deformation 
depends on an infinite number of free parameters.
For a certain choice of the bilinear form,
the renormalised circle product is identical with
the renormalised t-product.

We call {\sl{quantum field algebra}} the symmetric
Hopf algebra $S(V)$ equipped with a Laplace pairing
and the group of renormalisation parameters. This name
seems appropriate because such a structure enables us
to define all the quantities that can be calculated
from quantum field theory and compared
with experiment.

The purpose of this paper is threefold:
(i) to define an algebraic framework which simplifies the combinatorics
of quantum field theory, (ii) to extend Rota's Laplace pairing
so that more general combinatorial problems can be solved,
(iii) to extract as much algebra as possible from the quantum field
formalism, so that algebraists can contribute to quantum field
theory without having to read physics textbooks.

In the next section, we define the Laplace Hopf algebra
and the circle product, we prove that the circle product
is associative and that, for a symmetric Laplace pairing,
the circle product reproduces the time-ordered product
and satisfies Wick's theorem. Then, 
we introduce the renormalised circle product, we prove
that it is associative and we
show that it reproduces the renormalised time-ordered product.
Then we give some algebraic properties of the bare and
renormalised time-ordered products. Finally, we describe
further connections between our algebraic approach and
quantum field theory. An appendix gathers the basic
concepts of Hopf algebras.

In this article we concentrate on the presentation of
the formalism and we give several proofs in full detail,
to show how straightforward they become in the new algebraic framework.
To make things as simple as possible we treat here scalar
bosons fields, but the extension to fermions is possible.

\section{Laplace Hopf algebra}

As pointed out by Fauser \cite{Fauser}, 
Wick's theorem is related to a mathematical
structure discovered by Rota and Stein
and called a Laplace Hopf algebra \cite{RotaStein94}.
Thus, our first step will be to give a definition
and a few properties of Laplace Hopf algebras.
The readers who are not familiar with Hopf
algebras should first read the appendix.

\subsection{The symmetric Hopf algebra}
Since we are working with bosons, our starting
point will be the symmetric algebra. Later, the symmetric product
will designate the normal product of bosonic creation
and annihilation operators (see section \ref{normalprodsect}).

Let $V$ be a finite dimensional vector space, with basis
$\{e_i\}$. We define the symmetric algebra $S(V)$ as the
direct sum
\begin{eqnarray*}
S(V) &=& \bigoplus_{n=0}^\infty S^n(V),
\end{eqnarray*}
where $S^0(V)=\mathbb{C}$, $S^1(V)=V$
and $S^n(V)$ is spanned by elements of the
form $e_{i_1}\vee\cdots\vee e_{i_n}$,
with $i_1 \leq i_2 \leq\dots\leq i_n$.
The symbol $\vee$ denotes the associative and commutative product
$\vee : S^m(V) \otimes S^n(V) \longrightarrow S^{m+n}(V)$
defined on elements of the basis of $V$ by
$(e_{i_1}\vee\cdots\vee e_{i_m})\vee (e_{i_{m+1}}\vee\cdots\vee e_{i_{m+n}}) =
e_{i_{\sigma(1)}}\vee\cdots\vee e_{i_{\sigma(m+n)}}$,
where $\sigma$ is the permutation on $m+n$ elements such that
$i_{\sigma(1)}\leq\cdots\leq i_{\sigma(m+n)}$,
and then extended by linearity and associativity
to all elements of $S(V)$. The unit of $S(V)$ is the scalar unit
$1\in \mathbb{C}$ (i.e. for any $u\in S(V)$: $1\vee u=u\vee 1=u$).
Moreover, this algebra is graded: if $u\in S^n(V)$ then we say that
its grading is $|u|=n$. The symmetric product is a graded map:
if $|u|=m$ and $|v|=n$ then $|u\vee v|=m+n$.
In fact, $S(V)$ can be seen as the algebra of polynomials
in the variables $\{e_i\}$, the elements of
$S^n(V)$ being homogeneous polynomials of degree $n$.

In this paper, $u$, $v$, $w$ designate elements of $S(V)$,
$a$, $a_i$, $b$, $c$ designate elements of $V$ and
$e_i$ are elements of a basis of $V$.

We define a coproduct over $S(V)$ as follows
\begin{eqnarray}
\Delta 1 &=& 1\otimes 1,\nonumber\\
\Delta a &=& a\otimes 1 + 1\otimes a\quad{\mathrm{for}}\quad a\in V, \label{Deltaa}\\
\Delta (u\vee v)&=& \sum (u_{(1)}\vee v_{(1)}) \otimes (u_{(2)}\vee v_{(2)}),\label{Deltau}
\end{eqnarray}
where Sweedler's notation was used for the coproduct of $u$ and $v$:
$\Delta u=\sum u_{(1)}\otimes u_{(2)}$ and
$\Delta v=\sum v_{(1)}\otimes v_{(2)}$.
This coproduct is coassociative and cocommutative.
It is equivalent to the coproduct defined by Pinter
(equation (5) of reference \cite{PinterHopf}).
As an example, we calculate the coproduct $\Delta (a\vee b)$,
where $a$ and $b$ are in $V$.
From equation (\ref{Deltau})
\begin{eqnarray*}
\Delta (a\vee b)&=& \sum (a_{(1)}\vee b_{(1)}) \otimes (a_{(2)}\vee b_{(2)}),
\end{eqnarray*}
From the definition (\ref{Deltaa}) of the coproduct acting on elements of $V$ we know that
$\Delta a = a\otimes 1 + 1\otimes a$
and
$\Delta b = b\otimes 1 + 1\otimes b$.
Thus, 
\begin{eqnarray*}
\Delta (a\vee b)&=& \sum (a_{(1)}\vee b_{(1)}) \otimes (a_{(2)}\vee b_{(2)})\\
&=& (a\vee b)\otimes (1\vee 1) +
    (a \vee 1)\otimes (1\vee b) +
    (1 \vee b) \otimes (a \vee 1) 
\\&& + (1 \vee 1) \otimes (a \vee b)
\\&=&
(a\vee b)\otimes 1 + a \otimes  b + b \otimes a + 1 \otimes (a \vee b).
\end{eqnarray*}
At the next order, we have
\begin{eqnarray*}
\Delta (a\vee b\vee c) &=&
 1\otimes a\vee b\vee c
 +a \otimes b\vee c
 +b \otimes a\vee c 
 +c \otimes a\vee b
\\&&
 +a\vee b \otimes c
 +a\vee c \otimes b
 +b\vee c \otimes a
+a\vee b\vee c\otimes 1.
\end{eqnarray*}

Generally, if
$u=a_1\vee a_2\vee \dots \vee a_n$,
the coproduct of $u$ 
can be written explicitly as (\cite{Loday98} p.450)
\begin{eqnarray}
\Delta u &=& u\otimes 1 + 1 \otimes u
\nonumber\\&&+
\sum_{p=1}^{n-1} \sum_{\sigma}
a_{\sigma(1)}\vee\dots\vee a_{\sigma(p)}
\otimes a_{\sigma(p+1)}\vee\dots\vee a_{\sigma(n)},
\label{shuffle}
\end{eqnarray}
where $\sigma$ runs over the $(p,n-p)$-shuffles.
A $(p,n-p)$-shuffle is a permutation $\sigma$ of
$(1,\dots,n)$ such that
\begin{eqnarray*}
\sigma(1)<\sigma(2)<\dots <\sigma(p)
\,\, {\mathrm{and}}\,\,
\sigma(p+1)<\dots <\sigma(n).
\end{eqnarray*}

The counit is defined by $\epsilon(1)=1$ and
$\epsilon(u)=0$ if $u\in S^n(V)$ and $n>0$.
The antipode is defined by $\antip(u)=(-1)^n u$ if $u\in S^n(V)$.
In particular, $\antip(1)=1$. Since the symmetric product
is commutative, the antipode is an algebra morphism:
$\antip(u\vee v)=\antip(u) \vee \antip(v)$. 

The fact that this coproduct is coassociative
and that $\Delta$, $\epsilon$ and $S$ give $S(V)$ 
the structure of a cocommutative Hopf algebra is a classical result
(\cite{Loday98} p.450 and references therein).

This algebra is called the symmetric Hopf algebra.

\subsection{The Laplace pairing}
We define a {\sl{Laplace pairing}} on $S(V)$ as a
bilinear form $V\times V\rightarrow \mathbb{C}$,
that we denote by $(a|b)$, and which is extended to $S(V)$ by the following
recursions
\begin{eqnarray}
(u\vee v|w) &=& \sum (u|w_{(1)}) (v|w_{(2)}), \label{laplaceid1}\\
(u | v\vee w) &=& \sum (u_{(1)}|v) (u_{(2)}|w).\label{laplaceid2}
\end{eqnarray}
It is important to stress that we do not assume any special symmetry
for the bilinear form (i.e. it is a priori neither symmetric nor
antisymmetric). The pairing defined by Rota and coll. in \cite{Grosshans}
is more general, but for quantum field theory our restricted definition
is sufficient.

The name ``Laplace pairing'' comes from the fact that equations (\ref{laplaceid1})
and (\ref{laplaceid2}) are an elegant way of writing the Laplace identities
for determinants (in the case of fermion operators).
These identities express the determinant in terms
of minors (see Ref.\cite{Vein} p.26 and Ref.\cite{Muir}
p.93). They were derived by Laplace in 1772 \cite{Laplace}.

In reference \cite{Grosshans}, it is proved that these
recursions have the following unique solution:
If $u=a_1\vee a_2\vee \dots \vee a_k$ and
$v=b_1\vee b_2\vee \dots \vee b_n$, then
$(u|v)=0$ if $n\not=k$ and
$(u|v)=\perm(a_i|b_j)$ if $n=k$.
We recall that the {\sl{permanent}} of the matrix  $(a_i|b_j)$ is
\begin{eqnarray}
\perm(a_i|b_j) &=& \sum_{\sigma} (a_1|b_{\sigma(1)}) \cdots (a_k|b_{\sigma(k)}),
\label{permanent}
\end{eqnarray}
where the sum is over all permutations $\sigma$ of $(1,\dots,k)$.
The permanent is a kind of determinant where all signs are positive.
For instance
\begin{eqnarray*}
(a\vee b| c\vee d) &=& (a|c)(b|d) + (a|d)(b|c).
\end{eqnarray*}

In this paper, a {\sl{Laplace Hopf algebra}} is the symmetric Hopf algebra
equipped with a Laplace pairing.

\section{The circle product}
This section introduces the circle product. 
Its importance stems form the fact that the circle product
treats in one single stroke the operator product and
the time-ordered product, according to the definition
of the bilinear form $(a|b)$.
Following \cite{RotaStein94}, the {\sl{circle product}} is the operation
on $S(V)$ defined by\footnote{In reference \cite{RotaStein94},
Rota and Stein define the circle product from 
a Laplace pairing that is not necessarily
scalar. This more general Laplace
pairing must satisfy two additional identities (called (c) and (d))
which are not necessary when the pairing is scalar. In fact,
their additional conditions are not even true in our case
(take for example $w=a$, $w'=b$ and $w''=1$ in their equation (c)).
Thus, the present circle product is not strictly a particular case
of Rota and Stein's.
However, the main properties of their circle product remain true
with a scalar pairing.}
\begin{eqnarray}
u\circ v &=& \sum u_{(1)}\vee v_{(1)} \, (u_{(2)}|v_{(2)}).
\label{defcircle}
\end{eqnarray}
By cocommutativity of the coproduct on $S(V)$, this definition is
equivalent to
\begin{eqnarray*}
u\circ v &=& \sum  u_{(1)}\vee v_{(2)} \, (u_{(2)}|v_{(1)})
= \sum  u_{(2)}\vee v_{(1)} \, (u_{(1)}|v_{(2)})
\\&=&
\sum  (u_{(1)}|v_{(1)})\, u_{(2)}\vee v_{(2)}.
\end{eqnarray*}

A few examples might be useful
\begin{eqnarray*}
a\circ b &=& a\vee b + (a|b), \\
(a\vee b)\circ c &=& a\vee b \vee c +(a|c)b + (b|c) a, \\
a\circ ( b \vee c) &=& a\vee b \vee c +(a|c)b + (a|b) c, \\
a\circ b \circ c &=& a\vee b \vee c + (a|b)c +(a|c)b + (b|c) a, \\
a\circ b \circ c \circ d &=& a\vee b \vee c\vee d + 
(a|b)c\vee d +(a|c)b\vee d 
\\&&\hspace*{-18mm}
+ (b|c) a\vee d
+(a|d)b\vee c +(b|d)a\vee c 
+ (c|d) a\vee b 
\\&&\hspace*{-18mm}
+(a|b)(c|d) +(a|c)(b|d) + (b|c) (a|d), \\
(a\vee b) \circ (c \vee d) &=& 
a\vee b \vee c\vee d 
+(a|c)b\vee d + (b|c) a\vee d
\\&&\hspace*{-18mm}
+(a|d)b\vee c +(b|d)a\vee c 
+(a|c)(b|d) + (b|c) (a|d).
\end{eqnarray*}
We shall need some simple properties of the circle product
\begin{eqnarray*}
u\circ 1 &=& 1\circ u = u, \\
u\circ (v+w) &=& u\circ v+u\circ w, \\
u\circ (\lambda v) &=& (\lambda u)\circ v= \lambda (u\circ v).
\end{eqnarray*}

As an exercise, we prove that
\begin{eqnarray}
(u|v) &=& \epsilon(u\circ v). \label{circepsilon}
\end{eqnarray}
We apply the counit $\epsilon$ to definition (\ref{defcircle})
of the circle product:
\begin{eqnarray*}
\epsilon(u\circ v) &=& \sum \epsilon(u_{(1)}\vee v_{(1)}) (u_{(2)}|v_{(2)}).
\end{eqnarray*}
The counit is an algebra morphism, i.e.
$\epsilon(u_{(1)}\vee v_{(1)})=\epsilon(u_{(1)})\epsilon(v_{(1)})$.
Thus, by linearity of the Laplace pairing,
\begin{eqnarray*}
\epsilon(u\circ v) &=& \sum (\epsilon(u_{(1)})u_{(2)}|\epsilon(v_{(1)})v_{(2)}).
\end{eqnarray*}
But the counit property is precisely the identity
$\sum \epsilon(u_{(1)})u_{(2)} = u$. This proves equation (\ref{circepsilon}).

In the next section, we prove that the circle product is associative:
$u\circ (v\circ w) = (u\circ v)\circ w$.

\subsection{Proof of associativity}
We give a detailed proof of the associativity of the circle product, because
it is an important result (and because neither Fauser nor Rota and Stein provided it).
We first need a useful lemma:
\begin{eqnarray}
\Delta (u\circ v) &=& \sum (u_{(1)}\vee v_{(1)}) \otimes (u_{(2)}\circ v_{(2)})
     \label{Deltacirc} \\
&=& \sum (u_{(1)}\circ v_{(1)})\otimes (u_{(2)}\vee v_{(2)}).
     \label{Deltacirc2}
\end{eqnarray}
The proof is easy, we start from the definition of the circle product
(\ref{defcircle})  to write
\begin{eqnarray*}
\Delta (u\circ v) &=& \sum \Delta(u_{(1)}\vee v_{(1)}) (u_{(2)}| v_{(2)}).
\end{eqnarray*}
Now we use the definition (\ref{Deltau}) of the coproduct of $u_{(1)}\vee v_{(1)}$:
\begin{eqnarray*}
\Delta (u\circ v) &=& \sum (u_{(11)}\vee v_{(11)})\otimes (u_{(12)}\vee v_{(12)}) (u_{(2)}| v_{(2)}).
\end{eqnarray*}
The coassociativity of the coproduct of $u$ means that any triple 
$(u_{(11)},u_{(12)},u_{(2)})$ can be replaced by
$(u_{(1)},u_{(21)},u_{(22)})$. Therefore,
\begin{eqnarray*}
\Delta (u\circ v) &=& \sum (u_{(1)}\vee v_{(11)})\otimes (u_{(21)}\vee v_{(12)}) (u_{(22)}| v_{(2)}).
\end{eqnarray*}
We use now the coassociativity of the coproduct of $v$
\begin{eqnarray*}
\Delta (u\circ v) &=& \sum (u_{(1)}\vee v_{(1)})\otimes (u_{(21)}\vee v_{(21)}) (u_{(22)}| v_{(22)}),
\end{eqnarray*}
and the definition of the circle product brings
\begin{eqnarray*}
\Delta (u\circ v) &=& \sum (u_{(1)}\vee v_{(1)})\otimes (u_{(2)}\circ v_{(2)}).
\end{eqnarray*}
To obtain the other identity of the lemma, we start from 
$u\circ v=\sum (u_{(1)}| v_{(1)}) u_{(2)}\vee v_{(2)}$ or we use the
cocommutativity of the coproduct.

The second lemma is
\begin{eqnarray}
(u|v\circ w) &=& (u\circ v|w). \label{lemma2}
\end{eqnarray}
The proof is straightforward
\begin{eqnarray*}
(u|v\circ w) &=& \sum (u|v_{(1)}\vee w_{(1)}) (v_{(2)}| w_{(2)})
\\&=&
\sum (u_{(1)}|v_{(1)})(u_{(2)}| w_{(1)}) (v_{(2)}| w_{(2)})
\\&=&
\sum (u_{(1)}|v_{(1)})(u_{(2)}\vee v_{(2)}| w)
=(u\circ v|w).
\end{eqnarray*}
The first line is the definition of the circle product (\ref{defcircle}), the second line is
the expansion of the Laplace pairing (\ref{laplaceid2}), the third
one is the Laplace identity (\ref{laplaceid1}) and the 
last equality is again equation (\ref{defcircle}).

The associativity of the circle product follows immediately from
these two lemmas. From definition (\ref{defcircle}) and the
first lemma we obtain
\begin{eqnarray*}
u\circ(v\circ w) &=&
\sum u_{(1)}\vee (v\circ w)_{(1)}\,(u_{(2)}| (v\circ w)_{(2)})
\\&=&
 \sum u_{(1)}\vee (v_{(1)}\vee w_{(1)})\,(u_{(2)}|v_{(2)}\circ w_{(2)}).
\end{eqnarray*}
From the associativity of the symmetric product and the second lemma
we obtain
\begin{eqnarray*}
u\circ(v\circ w) &=&
\sum u_{(1)}\vee v_{(1)}\vee w_{(1)}\,(u_{(2)}\circ v_{(2)}| w_{(2)}).
\end{eqnarray*}
Now the first lemma enables us to rewrite this as
\begin{eqnarray*}
u\circ(v\circ w) &=&
\sum (u\circ v)_{(1)}\vee w_{(1)}\,((u\circ v)_{(2)}| w_{(2)})
= (u\circ v)\circ w.
\end{eqnarray*}

In comparison with the standard proofs of quantum field theory,
the combinatorial derivations have been replaced by
purely algebraic ones. This is the great advantage of the
Hopf algebraic approach to combinatorics advocated by Rota.

\subsection{Using the antipode}
In reference \cite{RotaStein94}, Rota and Stein use the
antipode of the symmetric Hopf algebra to write the
symmetric product in terms of the circle product.
The same holds in our case:
\begin{eqnarray*}
u\vee v &=& \sum (\antip(u_{(1)})|v_{(1)}) u_{(2)}\circ v_{(2)}
= \sum (u_{(1)}|\antip(v_{(1)})) u_{(2)}\circ v_{(2)}.
\label{symmetric-circle}
\end{eqnarray*}
The proof is simple:
\begin{eqnarray*}
\sum (\antip(u_{(1)})|v_{(1)}) u_{(2)}\circ v_{(2)} &=&
\sum (\antip(u_{(1)})|v_{(1)}) (u_{(21)}|v_{(21)}) u_{(22)}\vee v_{(22)}
\\&=&
\sum (\antip(u_{(11)})|v_{(11)}) (u_{(12)}|v_{(12)}) u_{(2)}\vee v_{(2)}
\\&=&
\sum (\antip(u_{(11)})\vee u_{(12)}|v_{(1)}) u_{(2)}\vee v_{(2)}
\\&=&
\sum \epsilon(u_{(1)})(1 |v_{(1)}) u_{(2)}\vee v_{(2)}
\\&=&
\sum \epsilon(u_{(1)})\epsilon(v_{(1)}) u_{(2)}\vee v_{(2)} = u\vee v.
\end{eqnarray*}
The last line was obtained because, from equation (\ref{circepsilon}),
$(1 |v_{(1)})=\epsilon(1\circ v_{(1)})=\epsilon(v_{(1)})$.
As for \cite{RotaStein94}, it is also possible to recover the Laplace
pairing from the circle product:
\begin{eqnarray}
(u | v) &=& \sum \antip(u_{(1)}\vee v_{(1)}) \vee (u_{(2)}\circ v_{(2)}).
\label{laplace-circle}
\end{eqnarray}
Again, the proof is straightforward:
\begin{eqnarray*}
\sum \antip(u_{(1)}\vee v_{(1)})\vee (u_{(2)}\circ v_{(2)})
&=&
\sum \antip(u_{(1)})\vee \antip(v_{(1)})\vee (u_{(2)}\circ v_{(2)})
\\&&\hspace*{-20mm}=
\sum \antip(u_{(1)})\vee \antip(v_{(1)}) \vee u_{(21)}\vee v_{(21)}(u_{(22)}|v_{(22)})
\\&&\hspace*{-20mm}=
\sum \antip(u_{(11)})\vee \antip(v_{(11)}) \vee u_{(12)}\vee v_{(12)}(u_{(2)}|v_{(2)})
\\&&\hspace*{-20mm}=
\sum \epsilon(u_{(1)})\epsilon(v_{(1)}) (u_{(2)}|v_{(2)}) =(u|v).
\end{eqnarray*}
Rota and Stein \cite{RotaStein94} also provide a sort of distributivity
formula for circle and symmetric products, which still holds:
\begin{eqnarray}
u \circ (v\vee w) &=& \sum (u_{(11)}\circ v)\vee (u_{(12)}\circ w) \vee\antip(u_{(2)}).
\label{distributivity}
\end{eqnarray}
For this result we need the following lemma
\begin{eqnarray*}
\sum (u_{(1)}\circ v) \vee s(u_{(2)}) &=& \sum (u|v_{(1)}) v_{(2)},
\end{eqnarray*}
which is proved easily
\begin{eqnarray*}
\sum (u_{(1)}\circ v) \vee s(u_{(2)}) &=& \sum (u_{(11)} | v_{(1)})
               u_{(12)} \vee v_{(2)} \vee s(u_{(2)})
\\&=&
\sum (u_{(1)} | v_{(1)}) u_{(21)} \vee v_{(2)} \vee s(u_{(22)})
\\&=&
\sum (u_{(1)} | v_{(1)}) \epsilon(u_{(2)}) v_{(2)}
=
\sum (u | v_{(1)}) v_{(2)}.
\end{eqnarray*}
Now, we can prove (\ref{distributivity}).
\begin{eqnarray*}
X &=& \sum (u_{(11)}\circ v)\vee (u_{(12)}\circ w) \vee\antip(u_{(2)}) 
\\&=&
\sum (u_{(1)}\circ v)\vee (u_{(21)}\circ w) \vee\antip(u_{(22)}) 
\\&=&
\sum (u_{(1)}\circ v)\vee w_{(2)} (u_{(2)}| w_{(1)}),
\end{eqnarray*}
where the last line was obtained using the previous lemma.
Now we rewrite the last line as
\begin{eqnarray*}
X &=& \sum (u_{(1)}| w_{(1)}) (u_{(2)}\circ v)\vee w_{(2)}
\\ &=&
\sum (u_{(1)}| w_{(1)}) (u_{(21)}| v_{(1)}) u_{(22)}\vee v_{(2)}\vee w_{(2)}
\\&=&
\sum (u_{(11)}| w_{(1)}) (u_{(12)}| v_{(1)}) u_{(2)}\vee v_{(2)}\vee w_{(2)}
\\&=&
\sum (u_{(1)}| w_{(1)} \vee v_{(1)}) u_{(2)}\vee v_{(2)}\vee w_{(2)}
=
u \circ (v\vee w).
\end{eqnarray*}

\subsection{Wick's theorem}
In this section, we show that 
the circle product satisfies a generalized version of 
Wick's theorem. This result is important because it
will be used to show that, according to the choice
of the bilinear form $(a|b)$, the circle product
is identical with the time-ordered product 
or with the operator product.

There are two Wick's theorems, one for operator products
and one for time-ordered products \cite{Gross}, but
they have an identical structure. Wick's theorem
is very well known, so we recall it briefly.
It states that the time-ordered (resp. operator)
product of a given number of elements of $V$
is equal to the sum over all possible pairs of
contractions (resp. pairings). 
The reader who is not familiar with Wick's
theorem will be referred to standard references
(e.g. \cite{Fetter} p.209, 
\cite{Weinberg} p.261, \cite{Ticciati} p.85).
A contraction $a^\bullet b^\bullet$
is the difference between the time-ordered
product and the normal product. In Wick's
notation \cite{Wick} 
$a^\bullet b^\bullet= T(ab)-{:}ab{:}$. 
A pairing $a^\diamond b^\diamond$
is the difference between the operator
product and the normal product:
$a^\diamond b^\diamond= ab-{:}ab{:}$. 
This pairing was used by Houriet and Kind
even before Wick's article \cite{Houriet}.
Both the contraction and the pairing are scalars.
To express these in our notation, we identify the
normal product ${:}ab{:}$ with the symmetric
product $a\vee b$. This is valid because
the normal product has all the properties
required for a symmetric product
(see section \ref{normalprodsect}).

On the one hand, the time-ordered product is symmetric, and
$a^\bullet b^\bullet$ is a symmetric bilinear
form that we can identify with our $(a|b)$.
Thus, the time-ordered product of two operators
is equal to the circle product obtained from
this symmetric bilinear form.

On the other hand, the pairing defined
from the operator product is an antisymmetric
bilinear form:
$a^\diamond b^\diamond= (ab-ba)/2$,
that we can also identify with our $(a|b)$
(not the same as in the previous paragraph, of course).
The operator product of two elements of $V$
is now equal to the circle product obtained
from this antisymmetric bilinear form.

Therefore, we have identified the symmetric products
with the normal products and we have shown that the circle product 
can reproduce the time-ordered
product and the operator product of two
elements of $V$. To prove that the equality
remains true for the product of more than two
elements of $V$, we shall show that the
circle product satisfies Wick's theorem.

We recall that the main ingredient used by Wick \cite{Wick}
to go from a product of $n$ elements of $V$ to a product
of $n+1$ elements of $V$ is the following recursive identity
\begin{eqnarray*}
{:}a_1\dots a_n{:}\,b &=& {:}a_1\dots a_n b{:}+
\sum_{j=1}^n a_j^\bullet b^\bullet\, {:}a_1\dots a_{j-1}a_{j+1}\dots a_n{:}
\end{eqnarray*}
This identity, the properties of the normal (i.e. symmetric)
product and the definition of the contraction (or pairing) of 
two operators were sufficient for Wick to prove his theorem.
Thus, we must show that the above identity is valid for
the circle product. In our notation, (defining $u=a_1\vee\dots\vee a_n$) 
we must prove
\begin{eqnarray}
u\circ b &=& u \vee b 
+ \sum_{j=1}^n (a_j|b)\,a_1\vee\dots\vee a_{j-1}\vee a_{j+1}\vee\dots\vee a_n.
\label{ucircb}
\end{eqnarray}
To show this, we use the definition
(\ref{defcircle}) of the circle product and equation (\ref{Deltaa}) to find
\begin{eqnarray*}
u\circ b &=& \sum u \vee b + \sum (u_{(1)}|b) u_{(2)}.
\end{eqnarray*}
The Laplace pairing $(u_{(1)}|b)$ is zero if the grading of $u_{(1)}$ is different from 1.
In other words, $u_{(1)}$ must be an element of $V$. 
According to the general equation (\ref{shuffle}) for $\Delta u$, this 
happens only for the $(1,n-1)$-shuffles.
By definition, a $(1,n-1)$-shuffle is a permutation 
$\sigma$ of $(1,\dots,n)$ such that
$\sigma(2)<\dots <\sigma(n)$, and the corresponding 
terms in the coproduct of $\Delta u$
are
\begin{eqnarray*}
\sum_{j=1}^n a_j \otimes a_1\vee\dots\vee a_{j-1}\vee a_{j+1}\vee\dots\vee a_n.
\end{eqnarray*}
This proves the required identity, and we have shown that the circle product
satisfies Wick's theorem.

Our proof of Wick's theorem for the circle product did not use any
symmetry of the bilinear form. Thus it is more general than the
usual Wick's theorem.  If the bilinear form is 
half of the commutator, as in Houriet and Kind's article \cite{Houriet},
then we obtain Wick's theorem for operator products
(\cite{Gross} p.212) and the circle product is equal to the
operator product.
If the bilinear form is 
the Feynman propagator, as in Wick's article \cite{Wick},
then we obtain Wick's theorem for time-ordered products
(\cite{Gross} p.215) and the circle product is equal to the
time-ordered product.
The fact that the commutator and the Feynman propagator
are sufficient to determine the operator and time-ordered
products was noticed by Fauser for the case of fermion operators
\cite{Fauser98}.

\section{Renormalisation}

In this section, we introduce the renormalised circle product
that we write $\barcirc$. In a finite dimensional vector space, the
purpose of renormalisation is no longer to remove infinities, since
everything is finite, but to provide a deformation of the circle product
involving an infinite number of parameters.
The physical meaning of these renormalisation parameters is the following:
time ordering of operators is clear when two operators are defined
at different times. However, the meaning of the time-ordering of 
two operators taken at the same time is ambiguous. Renormalisation
is here to parametrise this ambiguity.
The renormalised circle product will be defined in several steps.

\subsection{Renormalisation parameters}
We first need the renormalisation parameters.
They are defined as a linear map $\Z$ from $S(V)$ to $\mathbb{C}$,
such that $\Z(1)=1$ and $\Z(a)=0$ for $a\in V$.
These parameters form a group: the renormalisation group. The
group product $\star$ is defined by
\begin{eqnarray*}
(\Z\star \Z')(u) &=& \sum \Z(u_{(1)}) \Z'(u_{(2)}).
\end{eqnarray*}
This product is called the convolution of the Hopf algebra.
The coassociativity and cocommutativity of the Hopf
algebra implies that the product $\star$ is associative and
commutative.
The unit of the group is the counit $\epsilon$ of the Hopf algebra.
The inverse of $\Z$ is $\Z^{-1}$, defined by
\begin{eqnarray*}
(\Z\star \Z^{-1})(u) &=& \epsilon(u),
\end{eqnarray*}
or, recursively, by
\begin{eqnarray*}
\Z^{-1}(1) &=& 1,\\
\Z^{-1}(u) &=& -\Z(u) - {\sum}' \Z(u_{(1)}) \Z^{-1}(u_{(2)}),
\end{eqnarray*}
where $\sum' u_{(1)}\otimes u_{(2)} = \Delta u - 1\otimes u - u\otimes 1$ for $u\in S^n(V), n>0$.
For instance,
\begin{eqnarray*}
\Z^{-1}(a) &=& 0, \\
\Z^{-1}(a\vee b) &=& -\Z(a\vee b), \\
\Z^{-1}(a\vee b\vee c) &=& -\Z(a\vee b\vee c), \\
\Z^{-1}(a\vee b\vee c\vee d) &=& -\Z(a\vee b\vee c\vee d) + 2 \Z(a\vee b) \Z(c\vee d) 
\\&& + 2 \Z(a\vee c) \Z(b\vee d) + 2 \Z(a\vee d) \Z(b\vee c).
\end{eqnarray*}

\subsection{Z-pairing}
From the renormalisation parameters, we can define a pairing
that we call a Z-pairing
\begin{eqnarray}
\ZZ(u,v) &=& \sum \Z^{-1}(u_{(1)}) \Z^{-1}(v_{(1)}) \Z(u_{(2)}\vee v_{(2)}).
\label{Zcouplingdef}
\end{eqnarray}
A few examples and properties might be useful
\begin{eqnarray*}
\ZZ(u,v) &=& \ZZ(v,u),\\
\ZZ(1,u) &=& \epsilon(u),\\
\ZZ(a,b) &=& \Z(a\vee b),\\
\ZZ(a,b\vee c) &=& \Z(a\vee b\vee c),\\
\ZZ(a\vee b,c\vee d) &=& \Z(a\vee b\vee c\vee d)-\Z(a\vee b) \Z(c\vee d),\\
\ZZ(a, b\vee c \vee d) &=& \Z(a\vee b\vee c\vee d)-\Z(a\vee b) \Z(c\vee d)\\
  && -\Z(a\vee c) \Z(b\vee d)-\Z(b\vee c) \Z(a\vee d).
\end{eqnarray*}

The main property of the Z-pairing is the coupling identity
\begin{eqnarray}
\sum \ZZ(u_{(1)}\vee v_{(1)},w) \ZZ(u_{(2)} , v_{(2)})
&=&
\sum \ZZ(u,v_{(1)}\vee w_{(1)}) \ZZ(v_{(2)} , w_{(2)}).
\nonumber\\&&
\label{Zcouplingid}
\end{eqnarray}
For instance, the reader can check this identity for the
case $u=a$, $v=b$, $w=c\vee d$:
\begin{eqnarray*}
\ZZ(a\vee b,c\vee d) &=& \ZZ(a , b\vee c \vee d) +\ZZ(a , c) \ZZ(b , d)
+\ZZ(b , c) \ZZ(a , d).
\end{eqnarray*}

To show equation (\ref{Zcouplingid}), 
we first need an easy lemma
\begin{eqnarray*}
\sum \Z^{-1}(u_{(1)}\vee v_{(1)}) \ZZ(u_{(2)} , v_{(2)}) &=&  \Z^{-1}(u)  \Z^{-1}(v).
\end{eqnarray*}
We first use the definition of $\ZZ(u_{(2)} , v_{(2)})$, then
the cocommutativity of $u_{(2)}$ and $v_{(2)}$, then 
the coassociativity of $u$ and $v$, then the definition of
$\Z^{-1}(u_{(11)}\vee v_{(11)})$, and finally the definition
of the counit:
\begin{eqnarray*}
\sum \Z^{-1}(u_{(1)}\vee v_{(1)}) \ZZ(u_{(2)} , v_{(2)}) 
&=&  \sum \Z^{-1}(u_{(1)}\vee v_{(1)}) \Z^{-1}(u_{(21)}) 
\\&&\hspace*{10mm} \Z^{-1}(v_{(21)}) \Z(u_{(22)}\vee v_{(22)})
\\&&\hspace*{-40mm}=
\sum \Z^{-1}(u_{(1)}\vee v_{(1)}) \Z^{-1}(u_{(22)}) \Z^{-1}(v_{(22)}) \Z(u_{(21)}\vee v_{(21)})
\\&&\hspace*{-40mm}=
\sum \Z^{-1}(u_{(11)}\vee v_{(11)}) \Z(u_{(12)}\vee v_{(12)}) \Z^{-1}(u_{(2)}) \Z^{-1}(v_{(2)})
\\&&\hspace*{-40mm}=
\sum \epsilon(u_{(1)}\vee v_{(1)}) \Z^{-1}(u_{(2)}) \Z^{-1}(v_{(2)})
=  \Z^{-1}(u)  \Z^{-1}(v).
\end{eqnarray*}
To prove the coupling identity, we
expand the Z-pairing in terms of the renormalisation parameters
\begin{eqnarray*}
\sum \ZZ(u_{(1)}\vee v_{(1)},w) \ZZ(u_{(2)} , v_{(2)})
&=&\sum
\Z^{-1}(u_{(11)}\vee v_{(11)}) \Z^{-1}(w_{(1)}) 
\\&&\hspace*{-10mm}
\Z(u_{(12)}\vee v_{(12)}\vee w_{(2)})
\ZZ(u_{(2)} , v_{(2)}).
\end{eqnarray*}
Now we use the cocommutativity of the coproduct of $u_{(1)}$
and $v_{(1)}$ and the coassociativity of the coproduct of $u$ and $v$:
\begin{eqnarray*}
\sum \ZZ(u_{(1)}\vee v_{(1)},w) \ZZ(u_{(2)} , v_{(2)})
&=&                                                                                                                                      
\Z^{-1}(u_{(12)}\vee v_{(12)}) \Z^{-1}(w_{(1)})
\\&&\hspace*{-1mm}
\Z(u_{(11)}\vee v_{(11)}\vee w_{(2)})
\ZZ(u_{(2)} , v_{(2)})
\\&=&
\Z^{-1}(u_{(21)}\vee v_{(21)}) \Z^{-1}(w_{(1)})
\\&&\hspace*{-1mm}
\Z(u_{(1)}\vee v_{(1)}\vee w_{(2)})
\ZZ(u_{(22)} , v_{(22)}).
\end{eqnarray*}
Now we use the easy lemma to eliminate $\ZZ(u_{(22)} , v_{(22)})$
and the cocommutativity of the
coproduct of $w$ and we obtain
\begin{eqnarray*}
\sum \ZZ(u_{(1)}\vee v_{(1)},w) \ZZ(u_{(2)} , v_{(2)})
&=&
\Z^{-1}(u_{(1)}) \Z^{-1}(v_{(1)}) \Z^{-1}(w_{(1)})
\\&& \hspace*{5mm} 
\Z(u_{(2)}\vee v_{(2)}\vee w_{(2)}).
\end{eqnarray*} 
The right hand side of this equation is fully summetric
in $u$, $v$ and $w$. Therefore, all permutations
of $u$, $v$ and $w$ in the left hand side give the
same result. In particular, the permutation
$(u,v,w)\rightarrow (v,w,u)$ and the symmetry of $Z$ transform the left hand
side of equation (\ref{Zcouplingid}) into its right hand
side, which proves equation (\ref{Zcouplingid}).
The reader can check that the Laplace pairing also satisfies the
coupling identity:
\begin{eqnarray}
\sum (u_{(1)}\vee v_{(1)}|w) (u_{(2)} | v_{(2)})
&=&
\sum (u|v_{(1)}\vee w_{(1)}) (v_{(2)} | w_{(2)}).
\label{Lcouplingid}
\end{eqnarray}

For the renormalisation theory, it would be interesting to know the 
most general solution of the coupling identity. More precisely,
if $\ZZ$ is a bilinear map from $S(V)\times S(V)$ to $\mathbb{C}$
such that equation (\ref{Zcouplingid}) is satisfied,
$\ZZ(u,v)=\ZZ(v,u)$ and $\ZZ(1,u)=\epsilon(u)$, what is the
most general form of $\ZZ$?

\subsection{Modified Laplace pairing}
From the Z-pairing, we can define the modified Laplace pairing as
\begin{eqnarray}
\rlaplace{u}{v} &=& \sum \ZZ(u_{(1)},v_{(1)}) (u_{(2)} | v_{(2)}).
\label{mLcouplingdef}
\end{eqnarray}
A few examples and properties might be appropriate
\begin{eqnarray*}
\rlaplace{u}{1}  &=& \rlaplace{1}{u}  = \epsilon(u),\\
\rlaplace{a}{b}  &=& \Z(a\vee b) +(a|b),\\
\rlaplace{a}{b\vee c} &=& \Z(a\vee b\vee c),\\
\rlaplace{a\vee b\vee c}{d} &=& \ZZ(a\vee b\vee c , d) ,\\
\rlaplace{a\vee b}{c\vee d} &=& \ZZ(a\vee b,c\vee d) + (a\vee b|c\vee d) 
+ \ZZ(a,c) (b|d) 
\\&&+ \ZZ(a,d) (b|c) + \ZZ(b,c) (a|d) 
+ \ZZ(b,d) (a|c).
\end{eqnarray*}

The modified Laplace pairing satisfies also the coupling identity
\begin{eqnarray}
\sum \rlaplace{u_{(1)}\vee v_{(1)}}{w}\, \rlaplace{u_{(2)}}{v_{(2)}}
&=&
\sum 
\rlaplace{u}{v_{(1)}\vee w_{(1)}}\,
\rlaplace{v_{(2)} }{ w_{(2)}}.
\label{mLcouplingid}
\end{eqnarray}

We show this by expanding the definition of the modified Laplace pairing
\begin{eqnarray*}
Y &=& \sum \rlaplace{u_{(1)}\vee v_{(1)}}{w} \rlaplace{u_{(2)}}{v_{(2)}}
\\&=&
\sum \ZZ(u_{(11)}\vee v_{(11)},w_{(1)})
(u_{(12)}\vee v_{(12)}|w_{(2)}) \rlaplace{u_{(2)}}{v_{(2)}}
\\&=&
\sum \ZZ(u_{(1)}\vee v_{(1)},w_{(1)})
(u_{(21)}\vee v_{(21)}|w_{(2)}) \rlaplace{u_{(22)}}{v_{(22)}}
\\&=&
\sum \ZZ(u_{(1)}\vee v_{(1)},w_{(1)})
(u_{(22)}\vee v_{(22)}|w_{(2)}) \rlaplace{u_{(21)}}{v_{(21)}}
\\&=&
\sum \ZZ(u_{(11)}\vee v_{(11)},w_{(1)})
(u_{(2)}\vee v_{(2)}|w_{(2)}) \rlaplace{u_{(12)}}{v_{(12)}}.
\end{eqnarray*}
We use the definition of $\rlaplace{u_{(12)}}{v_{(12)}}$
\begin{eqnarray*}
Y &=& 
\sum \ZZ(u_{(11)}\vee v_{(11)},w_{(1)})
(u_{(2)}\vee v_{(2)}|w_{(2)}) 
\ZZ(u_{(121)},v_{(121)})
\\&&
(u_{(122)}|v_{(122)})
\\&=&
\sum \ZZ(u_{(111)}\vee v_{(111)},w_{(1)})
(u_{(2)}\vee v_{(2)}|w_{(2)}) 
\ZZ(u_{(112)},v_{(112)})
\\&&
(u_{(12)}|v_{(12)})
\\&=&
\sum \ZZ(u_{(11)}\vee v_{(11)},w_{(1)})
\ZZ(u_{(12)},v_{(12)})
(u_{(22)}\vee v_{(22)}|w_{(2)}) 
\\&&
(u_{(21)}|v_{(21)}).
\end{eqnarray*}
Now the same operations are done starting from
$\sum \rlaplace{u}{v_{(1)}\vee w_{(1)}} \rlaplace{v_{(2)} }{ w_{(2)}}$.
We must only do the permutation $(u,v,w)\rightarrow (v,w,u)$
and put $u$ on the left when $w$ was on the right.
This leads us to 
\begin{eqnarray*}
\sum 
\rlaplace{u}{v_{(1)}\vee w_{(1)}}\,
\rlaplace{v_{(2)} }{ w_{(2)}}
&=&
\sum \ZZ(u_{(1)} , v_{(11)} \vee w_{(11)})
\\&&\hspace*{-20mm}
\ZZ(v_{(12)},w_{(12)})
(u_{(2)}| v_{(22)} \vee w_{(22)}) 
(v_{(21)}|w_{(21)}).
\end{eqnarray*}
The right hand side of this equation is related to
the last form of Y by the 
coupling identities for
the Z-pairing and the Laplace pairing. 
Thus they are identical and 
the proof of equation (\ref{mLcouplingid})
is complete.

\subsection{Renormalised circle product}
After these preliminaries, we can now define the renormalised
circle product
\begin{eqnarray}
u \barcirc v &=& \sum \rlaplace{u_{(1)}}{v_{(1)}} \,u_{(2)} \vee v_{(2)}.
\label{rtproduitdef}
\end{eqnarray}
A few examples and properties might again be useful
\begin{eqnarray*}
1 \barcirc u &=& u \barcirc 1 = u,\\
\rlaplace{u}{v} &=& \epsilon(u \barcirc v),\\
a \barcirc b &=& a\vee b + \Z(a\vee b) +(a|b),\\
(a \vee b) \barcirc c &=& a\vee b\vee c +\Z(a\vee b\vee c)
+(a|c) b + (b|c) a 
\\&& + \Z(b\vee c) a + \Z(a\vee c) b.
\end{eqnarray*}

The renormalised circle product is associative.
To show that, we need the following identity
\begin{eqnarray}
\Delta (u\barcirc v) &=& \sum (u_{(1)}\vee v_{(1)}) \otimes (u_{(2)}\barcirc v_{(2)})
     \label{Deltabarcirc} \\
&=& \sum (u_{(1)}\barcirc v_{(1)})\otimes (u_{(2)}\vee v_{(2)}),
     \label{Deltabarcirc2}
\end{eqnarray}
which is derived exactly as equations (\ref{Deltacirc}) and (\ref{Deltacirc2}).
Therefore, using equation (\ref{Deltabarcirc}) and the cocommutativity and coassociativity of
the coproduct,
\begin{eqnarray*}
(u \barcirc v) \barcirc w &=& \sum \rlaplace{(u\barcirc v)_{(1)}}{w_{(1)}} 
(u\barcirc v)_{(2)}\vee w_{(2)}
\\&=&
\sum \rlaplace{u_{(1)}\vee v_{(1)}}{w_{(1)}} (u_{(2)}\barcirc v_{(2)}) \vee w_{(2)}
\\&=&
\sum \rlaplace{u_{(1)}\vee v_{(1)}}{w_{(1)}} \rlaplace{u_{(21)}}{v_{(21)}}
u_{(22)}\vee v_{(22)} \vee w_{(2)}
\\&=&
\sum \rlaplace{u_{(11)}\vee v_{(11)}}{w_{(1)}} \rlaplace{u_{(12)}}{v_{(12)}}
u_{(2)}\vee v_{(2)} \vee w_{(2)}.
\end{eqnarray*}
From the coupling identity (\ref{mLcouplingid}) we obtain
\begin{eqnarray*}
(u \barcirc v) \barcirc w &=& \sum 
\rlaplace{u_{(1)}}{v_{(11)}\vee w_{(11)}} 
\rlaplace{v_{(12)}}{w_{(12)}}
u_{(2)}\vee v_{(2)} \vee w_{(2)} 
\\&=&
\sum \rlaplace{u_{(1)}}{v_{(1)}\vee w_{(1)}} 
\rlaplace{v_{(21)}}{w_{(21)}}
u_{(2)}\vee v_{(22)} \vee w_{(22)} 
\\&=&
\sum \rlaplace{u_{(1)}}{(v\barcirc w)_{(1)}} 
u_{(2)}\vee (v_{(2)} \barcirc w_{(2)}) = u \barcirc (v\barcirc w).
\end{eqnarray*}
Thus, the renormalised circle product is associative. This is the
first main result of the paper.
Note that the action of the renormalisation group is not on the 
elements of the algebra but on the circle product.
The Laplace pairing and the renormalisation parameters provide
a deformation of the symmetric product with an infinite
number of parameters.
Now we shall define the t-product and the renormalised t-product, and
we shall prove that the renormalised circle product reproduces the
renormalised t-product of quantum field theory.

\section{The time-ordered product}
In quantum field theory, the boson operators commute inside a
t-product. Thus, from now on, the circle product will be considered commutative.
We first show that this is equivalent to $(a|b)=(b|a)$
for all $a,b\in V$. The corresponding result for Clifford algebras
was obtained by Fauser \cite{Fauser98}.

\subsection{Commutative circle product}
If the circle product is commutative, then
in particular $a\circ b=b\circ a$, so that $(a|b)=(b|a)$.
Inversely, if $(a|b)=(b|a)$
for all $a,b\in V$, then $u\circ v=v\circ u$ for all $u,v \in S(V)$.

The first step is to show that, if $(a|b)=(b|a)$
for all $a,b\in V$, then $(u|v)=(v|u)$ for all $u,v \in S(V)$.
But this follows immediately from the definition (\ref{permanent}) of
the permanent.
Because of the commutativity of the symmetric product we obtain
\begin{eqnarray*}
u\circ v &=& \sum u_{(1)}\vee v_{(1)} \, (u_{(2)}|v_{(2)})
=\sum v_{(1)}\vee u_{(1)} \, (v_{(2)}|u_{(2)}) = v \circ u.
\end{eqnarray*}

\subsection{The T-map}
We define recursively a linear map $T$ from $S(V)$ to $S(V)$ by
$T(1)=1$, $T(a)=a$ for $a$ in
$V$ and $T(u\vee v)=T(u)\circ T(v)$.
More explicitly, $T(a_1\vee\dots\vee a_n) = a_1\circ\dots\circ a_n$.
The circle product is associative and (in this section) commutative,
therefore $T$ is well defined.
This T-map is the usual t-product of quantum field theory.

The main property we shall need in the following is that 
\begin{eqnarray}
\Delta T(u) &=& \sum u_{(1)} \otimes T(u_{(2)})= \sum T(u_{(1)}) \otimes u_{(2)}.
\label{DeltaT}
\end{eqnarray}
We use a recursive argument. Equation (\ref{DeltaT}) is true
for $u=1$ or $u=a$ and $v=1$ or $v=b$.
Now
\begin{eqnarray*}
\Delta T(u\vee v) &=& \Delta \big( T(u)\circ T(v)\big).
\end{eqnarray*}
By equation (\ref{Deltacirc}) we obtain
\begin{eqnarray*}
\Delta T(u\vee v) &=& \sum \big(T(u)_{(1)}\vee T(v)_{(1)}\big)
\otimes
\big(T(u)_{(2)}\circ T(v)_{(2)}\big).
\end{eqnarray*}
The recursion hypothesis yields
\begin{eqnarray*}
\Delta T(u\vee v) &=& \sum \big(u_{(1)}\vee v_{(1)}\big)
\otimes
\big(T(u_{(2)})\circ T(v_{(2)})\big)
\\&=&
\sum (u\vee v)_{(1)}
\otimes
T(u_{(2)}\vee v_{(2)}).
\end{eqnarray*}
This gives us the expected result. The symmetric
identity in equation (\ref{DeltaT}) is obtained by
cocommutativity of the coproduct.

\subsection{Exponentiation}
Now, we shall derive a result obtained by Anderson \cite{Anderson}
and rediscovered several times \cite{Stumpf}:
the T-map can be written as the exponential of an operator $\Sigma$.
As an application of the Laplace Hopf algebra, we give a purely
algebraic proof of this.

First, we define a derivation $\delta_k$ attached to
a basis $\{e_k\}$ of vector space $V$. This derivation is defined
as the linear operator on $S(V)$ satisfying the following properties
\begin{eqnarray}
\delta_k 1 &=& 0,\nonumber\\
\delta_k e_j &=& \delta_{kj},\label{deriv1}\\
\delta_k (u\vee v) &=& (\delta_k u)\vee v + u \vee (\delta_k v).
\label{deriv2}
\end{eqnarray}
In the second equation, $\delta_{kj}$ is 1 for $j=k$ and
zero otherwise.
The Leibniz relation (\ref{deriv2}) gives us
$\delta_1 (e_1\vee e_2)=e_2$ and
$\delta_2 (e_1\vee e_2)=e_1$, by example.
From this definition, it can be shown recursively 
that the derivatives commute: $\delta_i\delta_j=\delta_j\delta_i$.
Note that the derivation does not act on the Laplace pairing:
$\delta_i (u|v) =0$.

Now we define the infinitesimal T-map as
\begin{eqnarray*}
\Sigma &=& \frac{1}{2} \sum_{ij} (e_i|e_j) \delta_i \delta_j,
\end{eqnarray*}
where the sum is over all elements of the basis of $V$.

We proceed in several steps to show that $T=\exp\Sigma$.
First, we show that 
\begin{eqnarray}
[\Sigma, a] &=& \sum_i (a|e_i) \delta_i. \label{Sigmaa}
\end{eqnarray}
To do that, we apply $\Sigma$ to an element $e_k\vee u$
\begin{eqnarray*}
\Sigma (e_k \vee u) &=& \frac{1}{2} \sum_{ij} (e_i|e_j) \delta_i \delta_j (e_k \vee u)
\\&=&
\frac{1}{2} \sum_{ij} (e_i|e_j) \delta_i (\delta_{jk} u + e_k \vee \delta_j u)
\\&=&
\frac{1}{2} \sum_{i} (e_i|e_k) \delta_i u
+\frac{1}{2} \sum_{j} (e_k|e_j) \delta_j u
+\frac{1}{2} \sum_{ij} (e_i|e_j) e_k \vee \delta_i \delta_j u
\\&=& \sum_{j} (e_k|e_j) \delta_j u +e_k \vee \Sigma u.
\end{eqnarray*}
If we extend this by linearity to $V$ we obtain
\begin{eqnarray}
\Sigma(a\vee u)=a\vee \Sigma u+\sum (a|e_j) \delta_j u. \label{Sigmaakwu}
\end{eqnarray}
And more generally
\begin{eqnarray*}
\Sigma (u \vee v) &=& (\Sigma u) \vee v + u \vee (\Sigma v)
+ \sum_{ij} (e_i|e_j) (\delta_i u )\vee (\delta_j v).
\end{eqnarray*}

From the commutation of the derivations we obtain
$[\Sigma,\delta_k]=0$ and $[\Sigma,[\Sigma, a]]=0$.
Therefore, the classical formula (\cite{Itzykson} p.167) yields
\begin{eqnarray*}
\ee^\Sigma a \ee^{-\Sigma} &=& a + [\Sigma,a],
\end{eqnarray*}
so that
\begin{eqnarray*}
[\ee^\Sigma,a] &=& \sum_i (a|e_i) \delta_i \ee^\Sigma = 
\ee^\Sigma \sum_i (a|e_i) \delta_i.
\end{eqnarray*}
This can be written more precisely as
\begin{eqnarray}
\ee^\Sigma (a\vee u) &=& a\vee (\ee^{\Sigma}u) +  [\Sigma,a] (\ee^{\Sigma}u). \label{eSigmaaveeu}
\end{eqnarray}

In the course of the proof of Wick's theorem, we derived
equation (\ref{ucircb}) which can be rewritten
\begin{eqnarray}
a\circ u &=& a\vee u + [\Sigma,a]u. \label{acircu}
\end{eqnarray}
Now we have all we need to prove inductively that $T=\ee^\Sigma$.
We have $T(1)=\ee^\Sigma 1=1$ and
$T(a)=a=\ee^\Sigma a$ because $\Sigma a=0$.
Now assume that the property is true up to grading $k$, we
take an element $u$ of $S^k(V)$ and calculate 
\begin{eqnarray*}
T(a\vee u) &=& a \circ T(u) = a\vee T(u) + [\Sigma,a] T(u)
\\&=& a \vee \ee^\Sigma u + [\Sigma,a] \ee^\Sigma u= \ee^\Sigma (a\vee u),
\end{eqnarray*}
where we used equation (\ref{acircu}), then equation (\ref{eSigmaaveeu}).
Thus the property is true for $a\vee u$ whose grading is $k+1$.

\subsection{The scalar t-map}
It is possible to write the T-map as a sum of scalars multiplied by
elements of $S(V)$. In fact we can show that
\begin{eqnarray}
T(u) &=& \sum t(u_{(1)}) u_{(2)}, \label{Tt}
\end{eqnarray}
where the map $t$ is a linear map from $S(V)$ to $\mathbb{C}$
defined recursively by
$t(1)=1$, $t(a)=0$ for $a\in V$ and
\begin{eqnarray}
t(u\vee v) &=& \sum t(u_{(1)}) t(v_{(1)}) (u_{(2)} | v_{(2)}). \label{deft}
\end{eqnarray}
This scalar map is called the t-map.
The proof is recursive. If the property is true
up to grading $k$, we take $w=u\vee v$, where $u$ and $v$ have
grading $k$ or smaller and we calculate
\begin{eqnarray*}
T(w) &=& T(u)\circ T(v) = \sum t(u_{(1)}) t(v_{(1)}) u_{(2)} \circ v_{(2)}
\\&=&
\sum t(u_{(1)}) t(v_{(1)}) (u_{(21)}|v_{(21)}) u_{(22)} \vee v_{(22)}
\\&=&
\sum t(u_{(11)}) t(v_{(11)}) (u_{(12)}|v_{(12)}) u_{(2)} \vee v_{(2)}
\\&=&
\sum t(u_{(1)}\vee v_{(1)}) u_{(2)} \vee v_{(2)} =\sum t(w_{(1)}) w_{(2)},
\end{eqnarray*}
where the first line is the definition of $T(u\vee v)$ and the 
recursion hypothesis, the second line is the definition of
the circle product, the third line is the coassociativity of the
coproduct and the last line is the definition of $t$.

Moreover, the t-map is well defined because
$t(u)=\epsilon(T(u))$. Again, this can be proved by recursion.
It is true for $u=1$ and $u=a$. If it is true up to grading
$k$, we take the same $w$ as for the previous proof and we have
\begin{eqnarray*}
\epsilon(T(w)) 
&=&
\sum t(w_{(1)}) \epsilon(w_{(2)})
=
t\big( \sum w_{(1)} \epsilon(w_{(2)})\big) = t(w).
\end{eqnarray*}

We close this section with a few examples:
$t(u)=0$ if $u\in S^n(V)$ with $n$ odd.
\begin{eqnarray*}
t(a \vee b) &=& (a|b),\\
t(a \vee b \vee c \vee d) &=& (a|b)(c|d) + (a|c)(b|d) +(a|d)(b|c).
\end{eqnarray*}
The general formula for $t(a_1 \vee \dots \vee a_{2n})$
has $(2n-1)!!$ terms which can be written
\begin{eqnarray*}
t(a_1 \vee \dots \vee a_{2n}) &=& 
\sum_\sigma \prod_{j=1}^{n}(a_{\sigma(j)}|a_{\sigma(j+n)}),
\end{eqnarray*}
where the sum is over the permutations $\sigma$
of $\{1,\dots,2n\}$ such that
$\sigma(1) < \sigma(2) < \cdots < \sigma(n)$
and $\sigma(j) < \sigma(n+j)$ for all $j=1,\dots,n$.
Or, as a sum over all the permutations $\sigma$
of $\{1,\dots,2n\}$.
\begin{eqnarray*}
t(a_1 \vee \dots \vee a_{2n}) &=& 
\frac{1}{2^n n!} \sum_\sigma (a_{\sigma(1)}|a_{\sigma(2)})
            \cdots (a_{\sigma(2n-1)}|a_{\sigma(2n)}).
\end{eqnarray*}

\section{The renormalised T-maps}
In this section, we define the renormalised T-maps,
and show that they coincide with the renormalised t-products of
quantum field theory.
A {\sl{renormalised T-map}} $\barT$ is a linear map
from $S(V)$ to $S(V)$ such that
$\barT(1)=1$, $\barT(a)=a$
for $a$ in $V$ and $\barT(u\vee v)=\barT(u)\barcirc \barT(v)$.
Since the circle product is assumed commutative, the renormalised circle
product is also commutative, and the map $\barT$ is well
defined on $S(V)$.

Following the proof of equation (\ref{DeltaT}), we can show that
\begin{eqnarray*}
\Delta \barT(u) &=& \sum u_{(1)} \otimes \barT(u_{(2)}).
\end{eqnarray*}

In this section we derive two renormalisation identities.
These identities show that our renormalised T-map coincides
with the renormalised t-product of quantum field theory,
as formalized by Pinter.

\subsection{First identity}
The first identity is:
\begin{eqnarray}
T(u) \barcirc T(v) &=& \sum \ZZ(u_{(1)} , v_{(1)}) T(u_{(2)})\circ T(v_{(2)}).
\label{bulltocirc}
\end{eqnarray}
This is proved as follows. By definition
\begin{eqnarray*}
T(u) \barcirc T(v) &=& \sum \rlaplace{T(u)_{(1)}}{T(v)_{(1)}} 
             T(u)_{(2)}\vee T(v)_{(2)}.
\end{eqnarray*}
Because of equation (\ref{DeltaT}), this becomes
\begin{eqnarray*}
T(u) \barcirc T(v) &=& \sum  \rlaplace{u_{(1)}}{v_{(1)}} 
   T(u_{(2)})\vee T(v_{(2)}).
\end{eqnarray*}
By definition of the modified Laplace pairing,
this can be rewritten
\begin{eqnarray*}
T(u) \barcirc T(v) &=& \sum \ZZ(u_{(11)} , v_{(11)})
   (u_{(12)} | v_{(12)})  
T(u_{(2)})\vee T(v_{(2)}),
\end{eqnarray*}
and the results follows from the coassociativity
of the coproduct and the definition of the circle product.

\subsection{Second identity}
Now we show the following formula
\begin{eqnarray}
\barT(u)&=& \sum \Z(u_{(1)}) T(u_{(2)}). \label{Pinter}
\end{eqnarray}
This is the second main result of the paper.
Equation (\ref{Pinter}) is Pinter's indentity for the 
renormalisation of a $t$-product (equation (17) in reference
\cite{PinterHopf}).
It is a very general renormalisation formula. Its importance
stems from the fact that it is valid also for field theories
that are only renormalisable through an infinite number of
renormalisation parameters. Its history starts with 
Dyson and Salam \cite{Dyson,Salam1,Salam2}.
A crypted version of it can be recognized with
hindsight in Bogoliubov's works \cite{BS56,BP57,Bogoliubov}.
It finally appeared in full light in Pinter's papers
\cite{Pinter,PinterHopf}.
Note that formula (\ref{Pinter}) does not
encompass the complete Epstein-Glaser renormalisation
\cite{Epstein,Scharf,Pinter} which completely avoids
infinities. It only describes the transition from one renormalisation
to another one. However, in the usual BPHZ renormalisation
of quantum field theory, the removal of infinities
is made by a formula which is equivalent to (\ref{Pinter}).
Thus, equation (\ref{Pinter}) is the standard BPHZ renormalisation
in the Epstein-Glaser guise. 

We are going to prove equation (\ref{Pinter}) recursively.
It is true for $u=1$ or $u=a$ and $v=1$ or $v=b$.
Now, suppose it holds for $u$ and $v$, then by definition,
and using the recursion hypothesis,
\begin{eqnarray*}
\barT(u\vee v) &=& \barT(u) \barcirc \barT(v) 
= \sum \Z(u_{(1)}) \Z(v_{(1)}) T(u_{(2)})\barcirc T(v_{(2)})
\end{eqnarray*}
Equation (\ref{bulltocirc}) gives us
\begin{eqnarray*}
\barT(u\vee v) 
&=& \sum \Z(u_{(1)}) \Z(v_{(1)}) \ZZ(u_{(21)}, v_{(21)})
T(u_{(22)})\circ T(v_{(22)}).
\end{eqnarray*}
From the coassociativity of the coproduct and the
definition (\ref{Zcouplingdef})
of the Z-pairing
we obtain
\begin{eqnarray*}
\barT(u\vee v) 
&=& \sum \Z(u_{(1)} \vee v_{(1)}) T(u_{(2)})
  \circ T(v_{(2)}) \\&=& \sum \Z((u\vee v)_{(1)}) T((u\vee v)_{(2)}).
\end{eqnarray*}
This is the required identity for $u\vee v$.

\subsection{The scalar renormalised t-map}
The reasoning that lead to the scalar t-map can
be followed exactly to define a scalar renormalised t-map
as a linear map $\bart$ from $S(V)$ to $\mathbb{C}$ such that
$\bart(1)=1$, $\bart(a)=0$ for $a\in V$ and
\begin{eqnarray}
\bart(u\vee v) &=& \sum \bart(u_{(1)}) \bart(v_{(1)}) 
  \rlaplace{u_{(2)}}{v_{(2)}}. \label{defbart}
\end{eqnarray}
Then $\bart(u)=\epsilon(\barT(u))$ and
\begin{eqnarray}
\barT(u) &=& \sum \bart(u_{(1)}) u_{(2)}, \label{barTt}
\end{eqnarray}
Moreover, equation (\ref{Pinter}) enables us to show
that 
\begin{eqnarray}
\bart(u) &=& \sum \Z(u_{(1)}) t(u_{(2)}).
\end{eqnarray}
We give a few examples
\begin{eqnarray*}
\bart(a \vee b) &=& (a|b) + \Z(a\vee b),\\
\bart(a \vee b \vee c) &=& \Z(a\vee b \vee c),\\
\bart(a \vee b \vee c \vee d) &=&  \Z(a\vee b \vee c \vee d)
+ \Z(a\vee b) (c|d) + \Z(a\vee c) (b|d) 
\\&&
+ \Z(a\vee d) (b|c) + \Z(b\vee c) (a|d) 
+ \Z(b\vee d) (a|c) 
\\&&
+ \Z(c\vee d) (a|b) 
+(a|b)(c|d) + (a|c)(b|d) +(a|d)(b|c).
\end{eqnarray*}

The scalar renormalised t-product corresponds to
the numerical distributions of the causal approach
\cite{Scharf,PinterHopf}.

\section{Relation to physics}
In this section, we make a closer connection
between the Hopf algebra approach and the usual
quantum field formalism.

\subsection{Normal products \label{normalprodsect}}
To define a normal product, 
we start from creation and annihilation operators
$a^+_k$ and $a^-_k$ that are taken as the basis vectors
of two vector spaces $V^{+}$ and $V^{-}$.
These two bases are in involution, i.e. there is an
operator $*$ such that ${a^+_k}^*=a^-_k$
and ${a^-_k}^*=a^+_k$.
The creation operators commute and there
is no other relation between them.
Thus, the Hopf algebra of the creation operators
is the symmetric algebra $S(V^+)$.
Similarly, the Hopf algebra of the annihilation
operators is $S(V^-)$.
From $V^{+}$ and $V^{-}$, we define a vector space $V=V^{+} \oplus V^{-}$,
so that each vector $a\in V$ can be written
$a=a^{+} + a^{-}$
and we call $P$ the projector from $V$ to $V^{+}$ and 
$M$ the projector from $V$ to $V^{-}$.
Thus $P(a)=a^+$ and $M(a)=a^-$.

There is an isomorphism between $S(V)$, the symmetric
algebra of $V$ and the tensor product of
the symmetric algebras $S(V^+)$ and $S(V^-)$.
$S(V)$ is the vector space of normal products.
The isomorphism $\varphi: S(V)\rightarrow S(V^+)\otimes S(V^-)$
is defined by $\varphi(u)=\sum P(u_{(1)})\otimes  M(u_{(2)})$
and we recover the fact that a normal product
puts all annihilation operators on the right of
all creation operators.
This isomorphism and the projectors 
$P : S(V)\rightarrow S(V^+)$ and
$M : S(V)\rightarrow S(V^-)$ can be defined
recursively by
\begin{eqnarray*}
P(1) &=& 1, \quad M(1) = 1, \\
P(a) &=& a^+, \quad M(a) = a^-, \\
\varphi(1) &=& 1\otimes 1 ,\\
\varphi(a) &=& P(a)\otimes 1 + 1\otimes M(a),\\
\varphi(u \vee v)  &=& \sum P(u_{(1)})P(v_{(1)}) \otimes  M(u_{(2)})M(v_{(2)}), \\
P(u \vee v) &=& P(u) P(v) , \quad M(u \vee v) = M(u) M(v).
\end{eqnarray*}
The algebra $S(V)$ is graded by
the number of creation and annihilation operators.

There is a subtelty here. 
It seems that we have forgotten the operator
product of elements of $V^{+}$ with elements
of $ V^{-}$. We have replaced $a^+_k a^{-}_l$
by $a^+_k \otimes a^{-}_l$ and we lost all information
concerning the commutation of creation and annihilation
operators. However, this is also true in standard quantum field
theory. Textbooks sometimes describe a kind of ``normal product
operator'' which takes a product of operators and
puts all creation operators on the left and all annihilation
operators on the right. But such an operator would
not be well-defined (\cite{Ticciati} p.28).
For instance, $a^{-}_k a^+_k$ and $a^+_k a^{-}_k +1$ 
are equal as operators, but their normal products
$a^+_k a^{-}_k$ and $a^+_k a^{-}_k +1$ are different.
Thus, it is not consistent to consider a normal product
as obtained from the transformation of an operator product
and we also loose all information concerning the
commutation relations in standard quantum field theory.

However, we saw that an antisymmetric Laplace pairing 
enables us to build a circle product in $S(V)$
which is an operator product. In that
sense normal products are a basic concept and 
operator product a derived concept of quantum field
theory. The problem of quantum field theory in
curved space times has also revealed that
normal products (i.e. Wick polynomials)
are fundamental elements of
quantum field theory \cite{Brunetti,Hollands}.
Therefore, from our point of view, quantum field
theory starts from the space $S(V)$ of normal products
and the operator product and time-ordered products
are obtained as deformations of the symmetric product.

In the quantum field theory of scalar particles,
our ``basis'' $a_k$  is not indexed by 
a finite number of $k$ but by a continuous
set of $x$. By an unfortunate convention,
the creation operators are
$\varphi^{(-)}(x)$
and the annihilation operators are
$\varphi^{(+)}(x)$.
Thus, for instance, the basis vectors of $V^+$ are the operators
$\varphi^{(-)}(x)$ where $x$ plays the role of the index of the
basis vectors.
This uncountable basis of $V^+$ is one of the sources of 
the analytical complexity of quantum field theory. 
For instance, these basic vectors enable us to 
define a bilinear form for
the definition of operator products as
(\cite{Itzykson} p.117)
\begin{eqnarray*}
(\varphi(x)|\varphi(y)) &=& i D(x-y),
\end{eqnarray*}
and a bilinear form for 
the definition of time-ordered products as
(\cite{Itzykson} p.124)
\begin{eqnarray*}
(\varphi(x)|\varphi(y)) &=& i G_F(x-y),
\end{eqnarray*}
where, for massless bosons (\cite{Scharf} p.93)
\begin{eqnarray*}
D(x) &=& - \frac{1}{2\pi} \sign(x^0) \delta(x^2),\\
G_F(x) &=& \frac{1}{4\pi^2} \frac{1}{x^2+i0},
\end{eqnarray*}
are distributions.
The manipulation of these distributions creates
many analytical problems that we do not want to address
here. This is why we consider only finite-dimensional
vector spaces in the present paper.

\subsection{The counit \label{counitsection}}
The counit is a basic element of Hopf algebras,
and it has a particularly simple expression for
quantum fields: it is the expectation value
over the vacuum.
In other words, for any element $u$ of $S(V)$
(i.e. for any sum of normal products of quantum fields)
the following striking identity holds:
$\epsilon(u)=\langle 0| u | 0\rangle$,
where $\epsilon$ is the counit of the symmetric
Hopf algebra.
This is very easy to show. Firstly,
$\langle 0| u | 0\rangle$ is a linear
map from $S(V)$ to $\mathbb{C}$. 
Secondly, for elements $u$ of $S^n(V)$ with $n>0$ 
$\epsilon(u)=0$ and $\langle 0| u | 0\rangle=0$
because $u$ is a normal product.
Finally, the symmetric Hopf algebra is connected, thus
all elements of $S^0(V)$ are multiple of
the unit: for any $u\in S^0(V)$ there is a
complex number $\lambda(u)$ such that
$u = \lambda(u) 1$.  Moreover, 
$\epsilon(u)=\lambda(u) \epsilon(1)=\lambda(u)$.
Thus $u = \epsilon(u) 1$.
But the vacuum is assumed to be normalized,
thus we have 
$\langle 0| u | 0\rangle=\epsilon(u)\langle 0| 1 | 0\rangle
=\epsilon(u)$. Thus the counit and the expectation value
over the vacuum are identical.
This identity is a additional argument in favor of the fact that
Hopf algebras are a natural framework for quantum field theory.

\subsection{The S-matrix and the Green function}
For any $u\in S(V)$ ($u$ plays the role of
a Lagrangian) we can define 
the normal ordered S-matrix $\Scal$ by
\begin{eqnarray*}
\Scal &=& \exp^\vee(u),
\end{eqnarray*}
where
\begin{eqnarray*}
\exp^\vee(u) &=& \sum_{n=0}^\infty \frac{1}{n!} u^{\vee n},
\end{eqnarray*}
and the symmetric power is defined recursively
by $u^{\vee 0}=1$, $u^{\vee 1}=u$ and
$u^{\vee (n+1)}=u\vee (u^{\vee n})$.
Now the bare S-matrix is defined as
$T(\Scal)$ and the renormalised S-matrix is
$\barT(\Scal)$. Of course, for a physical definition
of the renormalised S-matrix, we must inject various
renormalisation and causality conditions which determine the
renormalisation parameters.
The mathematically-oriented reader has noticed
that $\exp^\vee(u)$ is an infinite series that
does not belong to $S(V)$. There are two ways out
of this problem. The first solution is to put a topology
on $S(V)$ and to extend $S(V)$ so that infinite
series are admitted. If the circle product is still defined
on such an extended states, our approach enables us
to manipulate non perturbative quantites, such as
the S-matrix and the Green function (see below).
This is an advantage over the Feynman diagram approach.
The second solution is more conservative:
we can consider formal power series
such as $\exp^\vee(\lambda u)$. Such 
a formal power series is a compact way to represent
an infinite number of terms (i.e. one term for
each $\lambda^n$) \cite{Hollands3}.

In section \ref{counitsection} we saw that the
expectation value over the vacuum is equal to the
counit. Thus, the Gell-Man and Low definition of
the Green function (\cite{Itzykson} p.264) 
can be translated into the Hopf algebraic approach as:
the bare electron Green function $G_{ij}$
is defined by
\begin{eqnarray*}
G_{ij} &=& \frac{\epsilon(e_i\circ e_j\circ
T(\Scal))}{\epsilon(T(\Scal))},
\end{eqnarray*}
the renormalised electron Green function $\barG_{ij}$
is defined by
\begin{eqnarray*}
\barG_{ij} &=& \frac{\epsilon(e_i\circ e_j\circ
\barT(\Scal))}{\epsilon(\barT(\Scal))}.
\end{eqnarray*}
A Schwinger-Dyson equation can be derived for $G_{ij}$.

These examples were chosen to show that the
circle product allows for a compact 
definition of quantum field quantities.

\subsection{The simplest Lagrangians}

For the simplest Lagrangian $u=a\in V$
we obtain the following identity
\begin{eqnarray*}
T(\Scal) &=& \exp^\vee(a+(a|a)/2) = \exp((a|a)/2) \exp^\vee(a).
\end{eqnarray*}

A slightly more complicated Lagrangian is
\begin{eqnarray*}
u &=& \sum_i e_i \vee e_i.
\end{eqnarray*}
This Lagrangian is local in the sense that
there is no cross term $e_i \vee e_j$.
It corresponds to a mass term 
$\varphi(x)^2$ of quantum field theory.
To write the S-matrix we define the
matrix $M$ by its matrix elements $M_{ij}=(e_i|e_j)$.
Then
\begin{eqnarray*}
T(\Scal) &=& \frac{1}{\sqrt{\det(1-2M)}}
\exp^\vee\Big(\sum_{ij} e_i (1-2M)^{-1}_{ij} e_j \Big).
\end{eqnarray*}

Here, there is a small difference between the algebraic
and the quantum field approaches. To make $T$
a map from $S(V)$ to $S(V)$, we must replace
all symmetric products by circle products.
In particular, $T(\Scal)$ contains terms such
as 
\begin{eqnarray*}
T(u) &=& \sum_i e_i \circ e_i = 
\sum_i e_i \vee e_i + \sum_i (e_i|e_i).
\end{eqnarray*}
These terms $(e_i|e_i)$ are usually avoided
in quantum field theory because they
are infinite. 
From the physical point of view, we can say
that $u$ is a local Lagrangian and all the terms
of $u$ are taken at the same time. Therefore,
the time-ordered product inside $u$ is not well defined 
and we must choose a prescription for it that agrees
with experiment.
According to the Epstein-Glaser approach,
the correct time-ordered product
taken at the same time is determined by renormalisation.
In other words, the divergent term $(e_i|e_i)$
is removed by choosing 
a renormalisation parameter 
(for example $\Z(e_i \vee e_i)=-(e_i|e_i)$).
This corresponds to the removal of the density of 
occupied negative energy states in the Dirac sea picture
\cite{Dirac34}.
In fact, in quantum electrodynamics in an
external field, only a part of $(e_i|e_i)$ must
be removed and the polarisation charge 
$(e_i|e_i)+\Z(e_i \vee e_i)$ creates the 
Uehling potential which is observable \cite{Uehling,Mohr}.

Thus, for quantum electrodynamics, it is necessary to take
the circle product inside the Lagrangian $u$. This 
is natural from the algebraic point of view, and this agrees
with experiment.
However, for scalar fields this is not necessarily the case
and we need a way to keep normal products inside $T(u)$.
This is 
again provided by Rota and coll. \cite{Grosshans}, who define a
{\sl{divided power}} as the quantity $a^{(n)}=a^{\vee n}/n!$
for $a\in V$, with $a^{(0)}=1$ and $a^{(1)}=a$.
The divided powers have the following properties, which
are deduced from the properties of $a^{\vee n}$:
\begin{eqnarray*}
\Delta a^{(n)} &=& \sum_{k=0}^n a^{(k)}\otimes a^{(n-k)},
\end{eqnarray*}
$a^{(m)}\vee a^{(n)}=\binom{m+n}{m} a^{(m+n)}$,
$\antip(a^{(n)})= (-1)^n a^{(n)}$ and
\begin{eqnarray*}
(a^{(n)}| a_1\vee a_2\vee \dots\vee a_n) &=& (a|a_1)(a|a_2)\cdots(a|a_n).
\end{eqnarray*}
Thus
\begin{eqnarray*}
(a^{(n)}| b^{(n)}) &=& \frac{(a|b)^n}{n!} = (a|b)^{(n)}.
\end{eqnarray*}
The normal product is preserved inside $T(u)$ by considering the divided powers
as independent variables and by defining the T-map on them by
$T(a^{(n)})=a^{(n)}$. The use of divided powers and their relations
with the elements of $V$ are treated in detail in reference \cite{Grosshans}.

\section{Conclusion}
We hope that we have convinced the reader that
Hopf algebra is a powerful and natural tool
for quantum field theory. It enables us to transform combinatorial
reasoning into algebraic manipulations.

This paper was expository, and only the
simple case of scalar fermions was treated.
In a forthcoming publication we shall consider
the cases of fermions and more complex problems
such as the relation of the renormalization
parameters with the usual counteterms of the Lagrangian.
We shall also discuss the question of connected
products \cite{Dutsch}, which have a nice operadic interpretation
\cite{Getzler}.

Fermions are a relatively
straightforward extension of the present formalism.
The main trick is to twist the tensor product so as
to account for the anticommutation property of
fermions. Moreover, bosons and fermions can be
merged into a supersymmetric algebra \cite{Grosshans}.
All this is necessary to consider quantum electrodynamics.
The circle product of fermions is a generalization
of the Clifford algebra and of 
Hestenes' Geometric Algebra \cite{Hestenes,Crapo}.
By considering a more complicated twist, it
might be possible to work with quantum
groups \cite{Ilinski}
and braided quantum field theory \cite{Oeckl}.

Since our algebra is finite dimensional,
it is well suited to treat problems in 
lattice field theory.

The circle and renormalised circle products 
are deformations of the normal product, thus
they could be relevant to the deformation
theory of quantum fields \cite{Dito90,Dito92,Dutsch}
or to quantize Fedorov \cite{Dolgushev1} 
and symplectic \cite{Dolgushev2} manifolds.

But the main reason why the present formalism 
was developed is the calculation of the Green
function for a degenerate vacuum. Some
steps in this direction were done by Kutzelnigg
and Mukherjee \cite{Kutzelnigg}, but the combinatorial
formulas are very complex because they mix time-ordered
and commutation functions. We hope that our quantum
field algebra can be helpful for that problem.

\ack
I am extremely grateful to Alessandra Frabetti
for her constant help and for her many suggestions
and clarifications. This paper would have been
much worse without her contribution.
I am also very grateful to Bertfried Fauser for
the detailed e-mails he sent me to
explain his view of the connection between
Laplace Hopf algebra and Wick's theorem, and
for the many comments and corrections he made to the manuscript.
I thank Moulay Benameur for a very useful discussion.
This is IPGP contribution \#0000.

\appendix
\section*{Appendix}
\setcounter{section}{1}

In this appendix, we make a short presentation of
Hopf algebras.

To define a Hopf algebra, we first need an algebra.
An {\sl {algebra}} is a vector space $\Acal$ over $\mathbb{C}$ 
equipped with an associative linear product over $\Acal$,
denoted $\cdot$ and a unit, denoted $1$.
A product is {\sl {linear}} if for any $a, b, c \in \Acal$:
$a\cdot(b+c)=a\cdot b+ a\cdot c$,
$(a+b)\cdot c= a\cdot c + b \cdot c$ and
$a\cdot (\lambda b) = \lambda (a\cdot b) = (\lambda a)\cdot b$.
It is {\sl{associative}} if for any $a, b, c \in \Acal$:
$(a\cdot b)\cdot c= a\cdot (b\cdot c)$.
The {\sl{unit}} is the element of $\Acal$ such that, for any $a\in \Acal$,
we have $1\cdot a=a\cdot 1=a$.

As an example, we consider the algebra
$\Xcal$ generated by all sets of elements
$\{x_1,\dots,x_n\}$, where $x_i$ are vectors
in $\mathbb{R}^3$ for example.
The associative product is given by the union of sets
\begin{eqnarray*}
\{x_1,\dots,x_{m}\}\cdot \{y_1,\dots,y_{n}\}
&=& \{x_1,\dots,x_{m}, y_1,\dots,y_{n}\}.
\end{eqnarray*}
The unit 1 is the empty set.

The most unusual concept in a Hopf algebra is the
coproduct, denoted $\Delta$. 
In general, a {\sl{coproduct}} is a linear application
from $\Acal$ to $\Acal\otimes\Acal$, denoted $\Delta$.
The coproduct has various physical meanings.
Quite often, a coproduct 
can be considered as giving all the ways to split an element of $\Acal$
into two ``parts''.
For the example of $\Xcal$, the coproduct of
$\{x_1,\dots,x_n\}$ is defined as
\begin{eqnarray*}
\Delta \{x_1,\dots,x_n\}
&=& \sum_{I,I^c} I\otimes I^c,
\end{eqnarray*}
where the sum is over all subsets $I$ of 
$\{x_1,\dots,x_n\}$, and $I^c=\{x_1,\dots,x_n\} \backslash I$.
For instance,
\begin{eqnarray*}
\Delta 1 &=& 1\otimes 1, \\
\Delta \{x_1\} &=& \{x_1\} \otimes 1 + 1\otimes \{x_1\},\\
\Delta \{x_1,x_2\} &=& \{x_1,x_2\} \otimes 1 +  \{x_1\} \otimes\{x_2\}
+ \{x_2\} \otimes\{x_1\}\\&&+ 1\otimes \{x_1,x_2\},\\
\Delta \{x_1,x_2,x_3\} &=& \{x_1,x_2,x_3\} \otimes 1 
+  \{x_1\} \otimes\{x_2,x_3\}
+  \{x_2\} \otimes\{x_1,x_3\}
\\&&
+  \{x_3\} \otimes\{x_1,x_2\}
+ \{x_1,x_2\}\otimes\{x_3\}
+ \{x_1,x_3\}\otimes\{x_2\}
\\&&
+ \{x_2,x_3\}\otimes\{x_1\}
+ 1\otimes \{x_1,x_2,x_3\}.
\end{eqnarray*}

To recall this `splitting', the action of the coproduct is
denoted by $\Delta a = \sum a_{(1)} \otimes a_{(2)}$,
where $a_{(1)}$ and $a_{(2)}$ are the first and second ``parts''
of $a$.
For instance, if $a=\{x_1\} $, we have
$a_{(1)}=\{x_1\}$ and $a_{(2)}=1$ for the first
term of $\Delta a$ and 
$a_{(1)}=1$ and $a_{(2)}=\{x_1\}$ for its second term.
This is called Sweedler's notation.

In a Hopf algebra, the coproduct is compatible with
the product, in other words
$\Delta (a\cdot b) = (\Delta a)\cdot (\Delta b)$.
Or, more precisely
$\Delta (a\cdot b) = \sum (a_{(1)}\cdot b_{(1)}) \otimes (a_{(2)}\cdot b_{(2)})$.
In mathematical terms, we say that the coproduct is an {\sl{algebra homomorphism}}.

This compatibility is often used to define the coproduct. For instance, 
we can consider the algebra $\Jcal$ generated by the angular momentum operators
$J_x$, $J_y$ and $J_z$, where the product $\cdot$ is the operator product.
Then the coproduct of these operators is defined by
$\Delta J_i = J_i \otimes 1 + 1 \otimes J_i$. 
We recognize here the angular momentum operators acting on products
of states $|\psi\rangle_1 \otimes |\psi\rangle_2$, 
i.e. the coupling of angular momenta.  
This gives us the second physical meaning of the coproduct: it is a way
to go from a one-particle operator to a two-particle operator.
For products of operators $J_{i_1}\cdot \dots \cdot J_{i_n}$ the coproduct is
defined recursively by its compatibility with the operator product:
\begin{eqnarray*}
\Delta (J_{i_1}\cdot J_{i_2}\cdot \dots \cdot J_{i_n})
&=& \Delta (J_{i_1}) \cdot \Delta (J_{i_2}\cdot \dots \cdot J_{i_n}).
\end{eqnarray*}
For example
$\Delta (J_x \cdot J_y) = J_x \cdot J_y \otimes 1 + J_x\otimes J_y 
+ J_y \otimes J_x + 1 \otimes J_x \cdot J_y.$
Now it is natural to go from a two-particle operator to
a three-particle operator by repeating the action of the
coproduct. But since $\Delta a = \sum a_{(1)}\otimes a_{(2)}$
it is not clear on which side the second $\Delta$ should act:
on $a_{(1)}$ or on $a_{(2)}$? A very important property of 
Hopf algebras is that the result does not depend on which
side you apply the coproduct. 
For each $a_{(1)}$ we denote the action of the coproduct
by $\Delta a_{(1)} = a_{(11)}\otimes a_{(12)}$ and
for each $a_{(2)}$,  $\Delta a_{(2)} = a_{(21)}\otimes a_{(22)}$.
So, the fact that the action of the operator on three particles
does not depend on the order of the coproducts amounts to
\begin{eqnarray*}
(\Delta\otimes \Id) \Delta a &=& 
\sum (\Delta a_{(1)}) \otimes a_{(2)} =
\sum a_{(11)}\otimes a_{(12)} \otimes a_{(2)}
\\&=&
\sum a_{(1)}\otimes a_{(21)} \otimes a_{(22)}= 
\sum a_{(1)}\otimes (\Delta a_{(2)})
\\&=& (\Id\otimes \Delta) \Delta a.
\end{eqnarray*}
This property is called the {\sl {coassociativity}} of the coproduct.
By applying again the coproduct, we can define the action of the angular
momentum operators on many-particle states. The coassociativity ensures
that the result does not depend on the order used.
It can be checked that the coproduct of $\Xcal$ and $\Jcal$ are 
coassociative.

We still need two ingredients to make a Hopf algebra: a counit and
an antipode.
A {\sl{counit}} is a linear map from $\Acal$ to $\mathbb{C}$, denoted by $\epsilon$, such
that $\sum a_{(1)}\epsilon(a_{(2)}) = \sum\epsilon(a_{(1)}) a_{(2)}=a$.
In a Hopf algebra, the counit $\epsilon$ is an algebra homomorphism
(i.e. $\epsilon(a\cdot b)=\epsilon(a)\epsilon(b)$).
In most cases, the counit is very simple. For $\Xcal$ we have
$\epsilon(1)=1$,
$\epsilon(\{x_1,\dots,x_n\})=0$ for $n>0$, 
for $\Jcal$ we have $\epsilon(1)=1$,
$\epsilon(J_{i_1}\cdot\dots\cdot J_{i_n})=0$ for $n>0$.
Finally the {\sl{antipode}} is a linear map from $\Acal$ to $\Acal$ such that
$\sum a_{(1)}\cdot \antip(a_{(2)}) = \sum \antip(a_{(1)})\cdot a_{(2)}=\epsilon(a) 1$.
The antipode is a kind of inverse. Its main general property is
that $\antip(a\cdot b)= \antip(b) \cdot \antip(a)$.
In our examples: 
for $\Xcal$ the antipode is defined by
$\antip(1)=1$ and 
$\antip(\{x\})=-\{x\}$, so that 
$\antip(\{x_1,\dots,x_n\})=(-1)^n\{x_1,\dots,x_n\}$ for $n>0$,
for $\Jcal$ the antipode is defined by
$\antip(1)=1$ and
$\antip(J_{i})=-J_i$, so that 
$\antip(J_{i_1}\cdot\dots\cdot J_{i_n})=(-1)^n J_{i_n}\cdot\dots\cdot J_{i_1}$ for $n>0$.

To summarize a {\sl {Hopf algebra}} an algebra $\Acal$ equipped with
a coassociative coproduct $\Delta$, a counit $\epsilon$ and an antipode $\antip$,
such that $\Delta$ and $\epsilon$ are algebra homomorphisms.

A Hopf algebra is {\sl{commutative}} if, for any $a,b \in \Acal$:
$a\cdot b= b\cdot a$. A Hopf algebra is
{\sl{cocommutative}} if, for any $a\in \Acal$:
$\sum a_{(1)}\otimes a_{(2)}= \sum a_{(2)} \otimes a_{(1)}$,
in other words, if $a$ can be split into $a_{(1)}$ and
$a_{(2)}$, then there is also a splitting of $a$
where $a_{(2)}$ is the first part and $a_{(1)}$ is the second part.
The reader can check that $\Xcal$ and $\Jcal$ are cocommutative
Hopf algebras.

The last concept we shall use is that of a graded Hopf algebra.
An algebra is {\sl {graded}} if it can be written as the
vector sum of subsets
$\Acal={\Acal}_0 \oplus {\Acal}_1\oplus{\Acal}_2\oplus\cdots$.
If $a\in {\Acal}_m$, we denote the {\sl {grading}} of $a$ by
$|a|=m$. The grading must be compatible with the product:
if $a\in {\Acal}_m$ and $b\in {\Acal}_n$ then $(a\cdot b)\in {\Acal}_{m+n}$.
Therefore, $1\in {\Acal}_0$. If all elements of ${\Acal}_0$ 
can be written $\lambda 1$, the algebra is said to be {\sl {connected}}.

In our two examples, $\Xcal$ is a connected graded Hopf algebra, where the grading is
given by $|\{x_1,\dots,x_n\}|=n$, but $\Jcal$ is not a graded algebra:
If it were graded, we could consider that the grading of 
$J_i$ is 1. In that case, the grading of $J_x\cdot J_y$ and of
$J_y\cdot J_x$ would be 2. But the commutator identity
$J_x\cdot J_y-J_y \cdot J_x=iJ_z$ tells us that the grading of $J_x \cdot J_y-J_y \cdot J_x$
would be one, which is incompatible with the fact that 
the grading of $J_x \cdot J_y$ and of $J_y \cdot J_x$ are 2.

For more information on Hopf algebras, see Refs.\cite{Chari,Loday98}.

\end{document}